\documentclass[prl,twocolumn,showpacs,superscriptaddress]{revtex4-1}

\usepackage{graphicx}
\usepackage{dcolumn}
\usepackage{amsmath}
\usepackage{epsfig}
\usepackage{bm}
\usepackage{amssymb}
\usepackage{float}
\usepackage{natbib}
\usepackage{color}
\usepackage{hyperref}
\hypersetup{colorlinks=true,linkcolor=blue,citecolor=blue,urlcolor=black}
\usepackage{xspace}
\usepackage{array}

\usepackage[squaren]{SIunits}

\newcommand{\ket}[1]{\ensuremath{{\left|#1\right\rangle}}\xspace}

\newcommand{\urf}{r\hspace{-0.5mm}f}

\newcommand{\opv}[1]{\hat{\bm #1}} 

\begin{document}

\title{Coherent transfer between low-angular momentum and circular Rydberg states}

\author{A. Signoles}
\altaffiliation[Present address: ]{Physikalisches Instit\"ut, Universit\"at Heidelberg, Im Neuenheimer Feld 226, 69120, Heidelberg, Germany.}
\affiliation{Laboratoire Kastler Brossel, Coll{\`e}ge de France, CNRS, ENS-PSL Research University, UPMC-Sorbonne Universit{\'e}s, 11, place Marcelin Berthelot, 75231 Paris Cedex 05, France}
\author{E.K. Dietsche}
\author{A. Facon}
\author{D. Grosso}
\author{S. Haroche}
\author{J.M. Raimond}
\author{M. Brune}
\author{S. Gleyzes}
\email{gleyzes@lkb.ens.fr}
\affiliation{Laboratoire Kastler Brossel, Coll{\`e}ge de France, CNRS, ENS-PSL Research University, UPMC-Sorbonne Universit{\'e}s, 11, place Marcelin Berthelot, 75231 Paris Cedex 05, France}

\date{\today}


\begin{abstract}
We realize a coherent transfer between a laser-accessible low-angular-momentum Rydberg state and the circular Rydberg level with maximal angular momentum. This transfer is induced by a radiofrequency field with a high-purity $\sigma^+$ polarization tuned at resonance on Stark transitions inside the hydrogenic Rydberg manifold.
We observe over a few microseconds more than twenty coherent Rabi oscillations between the initial Rydberg state and the circular Rydberg level. We characterize in details these complex oscillations involving many Rydberg levels and find them to be in perfect agreement with a simple theoretical model. This coherent transfer procedure opens the way to hybrid quantum gates bridging the gap between optical communication and quantum information manipulations based on microwave Cavity and Circuit Quantum Electrodynamics.

\end{abstract}

\maketitle

The long-lived Circular Rydberg Levels (CRLs) are ideal tools for quantum manipulation of microwave (mw) fields stored in ultra-high-Q 3D superconducting resonators. They led to early demonstrations of basic quantum information processing operations~\cite{Raimond2001a} and to the generation of non-classical field state superpositions~\cite{Brune1996}. More recently, they have been instrumental in the exploration of fundamental Cavity Quantum Electrodynamics (Cavity-QED) effects, such as QND measurements of the photon number~\cite{Guerlin2007} and quantum feedback~\cite{Sayrin2011}. Recent advances on the manipulation of Rydberg atoms near atom chips~\cite{Teixeira2015} indicate that they could also be interfaced with the resonant structures used in the flourishing field of Circuit-QED~\cite{Hogan2012,Devoret2013}.

However, the photons used in mw Cavity- and Circuit-QED are unable to propagate over long-range transmission lines~\cite{Wang2011}. Optical to mw interfaces are thus the focus of an intense activity~\cite{Maxwell2014,Andrews2014}. A new realm for mw quantum information manipulation would open if the CRLs could be coherently interfaced with optical photons, which are ideal quantum information carriers over fiber and free-space communication networks~\cite{Cirac2017}. Unfortunately, the CRLs, with their large orbital quantum number $\ell= n-1$ ($n$ is the principal quantum number), do not couple directly to optical photons. 

In contrast, low-angular momentum Rydberg states are accessible from the ground state by coherent laser excitation~\cite{Johnson2008,Reetz2008,Miroshnychenko2010}. They were recently used for optical quantum information processing operations such as photon-photon gates relying on single-photon optical non-linearities induced by the dipole blockade mechanism~\cite{Peyronel2012,Parigi2012,Baur2014,Paredes2014}. They could also lead to quantum gates entangling a mw photon with a collective hyperfine excitation in a ground state atomic ensemble~\cite{Pritchard2014} and hence, through the DLCZ protocol~\cite{Duan2001}, to gates entangling optical and mw photons. However, the short lifetime of these levels, of the order of 100 $\mu$s, sets limits on the quantum transfer fidelity and on their use in CQED experiments. 

The missing link between mw and optical photons is a fast coherent transfer from a laser-accessible low-$\ell$ state to the CRLs. The most efficient CRL preparation technique so far involves a series of radiofrequency (rf) transitions between Stark levels performed in an rapid adiabatic passage~\cite{Hulet1983,Nussenzveig1993}. It thus requires a relatively long time, much longer than the typical Rabi frequency on the rf transitions. Therefore, the adiabatic rapid passage results in the accumulation of large dynamic phases. Their unavoidable experimental fluctuations affect the coherence of the process. The crossed fields CRLs preparation technique~\cite{Delande1988,Hare1988}, also based on an adiabatic passage, suffers from similar limitations.

In this Letter we demonstrate a fast, coherent transfer between a low-$\ell$ state and the CRL. By coupling a Rydberg atom to a rf field with a well-defined polarization, we isolate in the Stark manifold with the principal quantum number $n$, a subset of states behaving as a large angular momentum $J \sim n/2$. We resonantly drive this angular momentum and observe more than 20 coherent Rabi oscillations between the lowest (low-$m$) and uppermost (circular) energy states. 

\begin{figure}[t]
\centering
\includegraphics[width=0.45 \textwidth]{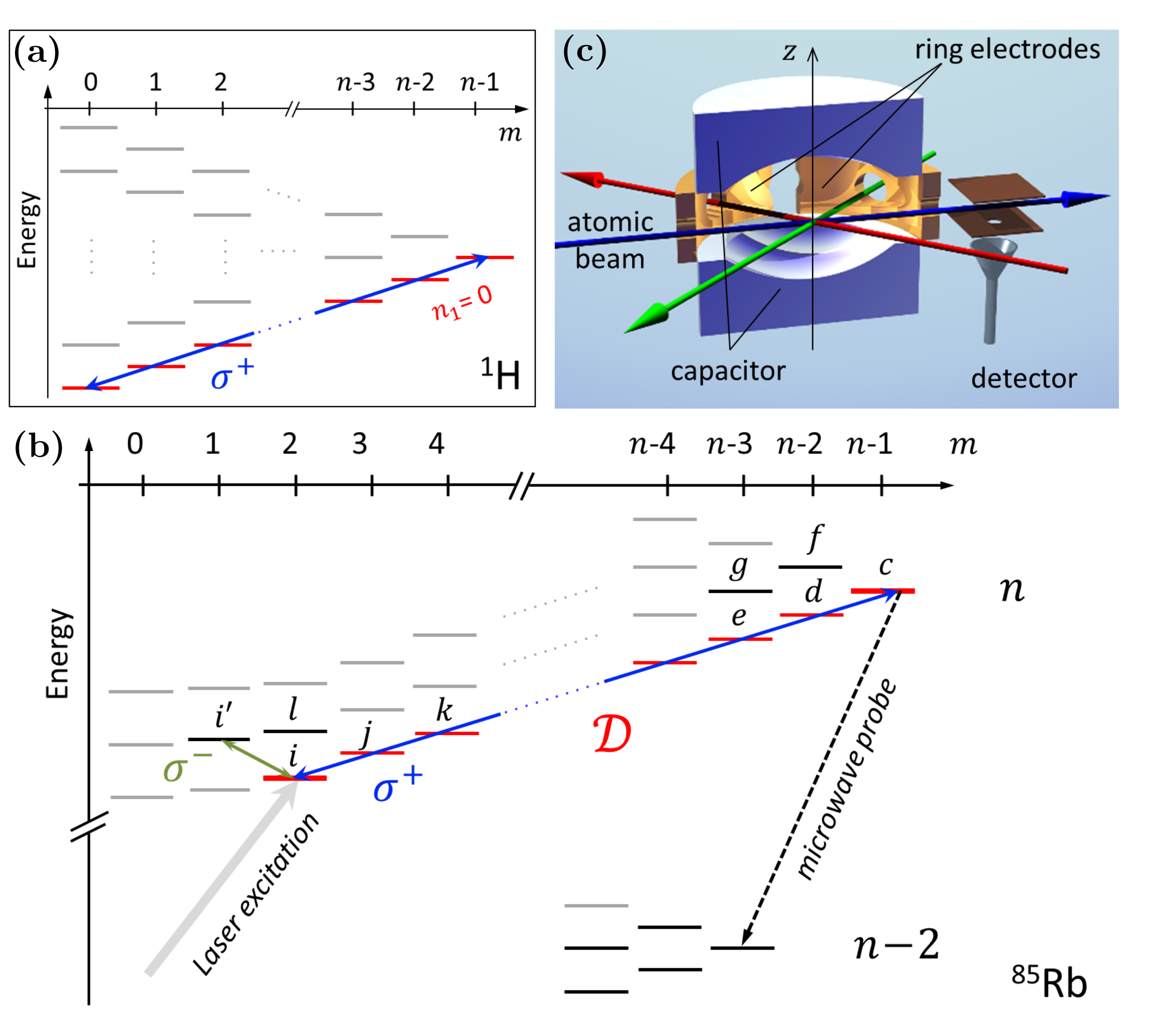}
\caption{(color online) (a) Scheme of the Hydrogen Rydberg levels in the $n$-th manifold in the presence of a static electric field (only states with $m\ge 0$ are shown). A $\sigma^+$ driving field couples all the $n_1=0$ levels (in red) from $m=0$ to the CRL. (b) Scheme of Rb Rydberg levels. A $\sim 2\,\volt\per\centi\meter$ electric field lifts the degeneracy of the $n$-th Stark manifold. States with $m \ge 3$ are hydrogenoid while  lower $m$ levels are affected by quantum defects. The ground state atoms are excited to the lower $m=2$ state of the manifold. A $\sigma^+$ polarized rf field couples \ket{n,i} to an angular momentum-like subspace $\cal D$ (red levels) via $\Delta  m=+1$ transitions, leading to a transfer toward the CRL \ket{n,c} (blue arrow). The population of each Stark level can be probed by selective mw transitions towards corresponding level in the $n-2$ manifold (dotted arrow). (c) Simplified experimental setup. Atoms from a thermal beam are optically excited at the center of the structure by laser beams (red: 780 nm and 776 nm  lasers and green : 1258 nm laser). A set of electrodes is used to apply static and rf electric fields. After a $\sim 230\,\micro\second$ time-of-flight the atoms are detected outside the structure by field ionization.}

\label{fig:scheme}
\end{figure}

The experiment is performed on $^{85}$Rb atoms in a static electric field $\bm F =F_0 \bm z$, which partly lifts the degeneracy of the Rydberg manifold. Its levels can be sorted by their magnetic quantum number $m$ (restricted to positive values here)~\cite{Gallagher1994}. For a Hydrogen atom (see Fig. \ref{fig:scheme}a), the Stark eigenstates are the parabolic states \ket{n,n_1,m}, where $0\le n_1\le n-m-1$ labels the states with the same $m$ from bottom to top in a regularly spaced ladder. The lowest level of each $m$ ladder ($n_1=0$) is midway between the lowest ones in the $m-1$ ladder. The $\sigma^+$ rf transitions between these lowest levels are thus all at the same angular frequency $\omega_n = 3/2\cdot nF_0 e a_0/\hbar$~\cite{Bethe1977} ($e$: elementary charge, $a_0$: Bohr radius; $\omega_n/2\pi\approx 100$ MHz for $F_0=1$ V/cm and $n\approx 50$). A $\sigma^+$ rf field couples all the $n_1=0$ states from $m=0$ up to the CRL. The subspace spanned by these levels is that of an angular momentum $\opv J$, with $J=(n-1)/2$~\cite{Lutwak1997}. In these terms, the CRL is the state \ket{J,m_J=J} (``angular momentum up", north pole on a generalized Bloch sphere) and the lowest-$m$ state \ket{n,0,0} is  \ket{J,m_J=-J} (``angular momentum down", south pole)~\cite{Englefield1972,Gay1989}. A resonant rf-drive at angular frequency $\omega_{\urf}=\omega_n$  induces a rotation of  $\opv J$ between the north and south poles at a frequency $\Omega_{\urf}/2\pi= 3/2nea_0 E^+_{\urf}/h$, where $E^+_{\urf}$ is the $\sigma^+$ polarized rf electric field amplitude. This rotation maps the angular momentum down state into the CRL in a time $\pi/\Omega_{\urf}$.

For Rubidium (Fig. \ref{fig:scheme}b), the levels with $m<3$ are shifted by the quantum defects due to the finite size of the atomic core~\cite{Gallagher1994}. Only the levels with $m\ge 3$ arrange in the hydrogenic pattern and are connected by degenerate $\sigma^+$ transitions (blue arrow on Fig. \ref{fig:scheme}b). For $m=2$, the $\ket{n,n_1=0,m=2} $ state is shifted far away from the hydrogenic multiplicity. However, a fortunate coincidence makes the rf transition between the laser-accessible  $\ket{n,n_1=1,m=2}=\ket{n,i}$ state and the first hydrogenic \ket{n,n_1=0,m=3}=\ket{n,j} level nearly degenerate with all other transitions toward $\ket{n,c}$. Hence, only the two lowest states $m=0,1$ are non-resonant. The transfer from $\ket{n,i}$ to the CRL is  a nearly $\pi$-rotation of the hydrogenic angular momentum $\opv J$, through a subset $\cal D$ of its levels (in red on Fig. \ref{fig:scheme}b).

The experiment takes place in a structure made up of a capacitor producing $\bm F$, surrounded by four ring electrodes, on which we apply the  rf signal at $\omega_{\urf}/2\pi =230$ MHz and dc potentials (Fig. \ref{fig:scheme}c). It is crossed by a Rubidium thermal beam (atomic velocity 250 m/s). Atoms are excited from the $5S$ ground state by three laser beams resonant on the transitions  $5S_{1/2}\rightarrow 5P_{3/2}$ (780 nm), $5P_{3/2}\rightarrow 5D_{5/2}$ (776 nm) and $5D_{5/2}\rightarrow nF,m=2$ (1258 nm). The 780 nm and 776 nm laser beams are $\sigma^+$-polarized and collinear with the quantization axis, defined by a dc electric field ($0.23\,\volt\per\centi\meter$) applied across the ring electrodes during the $1\,\micro\second$ pulsed laser excitation. The 1258 nm laser beam is $\pi$-polarized and perpendicular to the others. After laser excitation, a potential applied to the capacitor is switched-on in $1\ \mu$s, resulting in a field $\bm F = F_0 \bm z$ with $F_0\sim 2$ V/cm. This Stark switching prepares the initial state $\ket{n,i}$~\cite{Nussenzveig1993}.

At the end of the sequence, the atoms drift out of the capacitor towards a field-ionization detector, which resolves the initial and circular Rydberg states but doesn't resolve them from levels with similar energies. In order to selectively detect neighboring levels $\ket{n,p}$ with  $p=i,i',j,k,l,c,d,e,f,g$ (Fig. \ref{fig:scheme}b) we use selective mw probe pulses, which transfer $\ket{n,p}$ into  another manifold, easily resolved by the field ionization detector.

\begin{figure}[t]
\centering
\includegraphics[width=0.45 \textwidth]{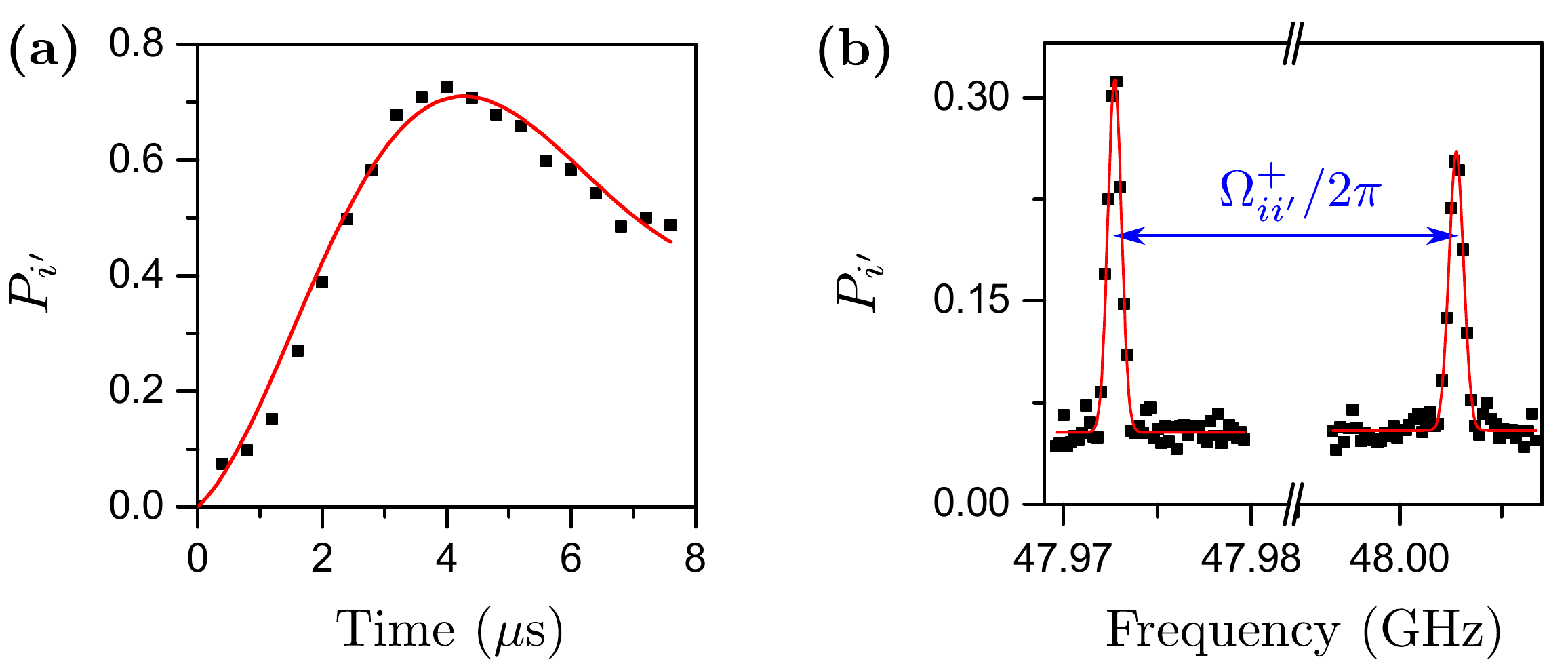}
\caption{(color online) (a) Rabi oscillation induced by the $\sigma^-$ polarization component of the rf field on the $\ket{52,i}\rightarrow\ket{52,i'}$ transition. Population $P_{i'}$ of $\ket{52,i'}$ versus the rf pulse duration $t$ (dots). We fit the frequency of the oscillation (red line) to $\Omega_{ii'}^-/2\pi = 107\pm25\,\kilo\hertz$. (b) Autler-Townes doublet of the $\ket{52,i}\rightarrow\ket{52,i'}$ transition dressed by the $\sigma^+$ rf field. Points are experimental and the red lines are Gaussian fits leading to $\Omega_{ii'}^+/2\pi=  30.0\,\mega\hertz$.}
\label{fig:polarization}
\end{figure}

The rf transfer from \ket{n,i} to \ket{n,c} requires a pure rf $\sigma^+$ polarization, in order to avoid transitions with $\Delta m=-1$ and to confine the evolution inside $\cal D$. By applying on the four ring electrodes 230 MHz signals with the proper phases and amplitudes, we can in principle generate a pure $\sigma^+$ field in the center of the structure. 

Due to the difference in the electrode driving line transmission and the capacitive coupling between the electrodes, these phases and amplitudes cannot be predetermined. We therefore use an optimization procedure, based on the direct measurement of the $\sigma^+$ and $\sigma^-$ amplitude. The $\sigma^-$ rf component is measured by tuning the electric field to $F_0=1.76$~V/cm, so that the transition between  \ket{n=52,n_1=1,m=2} and \ket{n=52,n_1=3,m=1} (\ket {52,i} and \ket {52,i'} on figure \ref{fig:scheme}b), coupled to the $\sigma^-$ field, is resonant at 230 MHz. All other transitions from $\ket {52,i}$ or $\ket {52,i'}$ are then out of resonance due to the quantum defects. The rf thus induces Rabi oscillation at a frequency $\Omega_{ii'}^-/2\pi=\sqrt 2 d_{ii'}E_{\urf}^-/h$ where $E_{\urf}^-$ is the amplitude of the $\sigma^-$ component and $d_{ii'}$ the dipole matrix element of this transition. In order to measure the  $\sigma^+$ field amplitude, we align the electric field along the $-\bm z$ direction instead of $+\bm z$. The atom is then prepared in the $m=+2$ state with respect to the reversed $Oz$ axis (noted $-Oz$). The transition between $\ket {52,i}$ and $\ket{52,i'}$ is driven by the $\sigma^-$ polarization w.r.t. $-Oz$, i.e. the $\sigma^+$ polarization w.r.t. $Oz$. 

The full optimization procedure is described in \cite{SI}. Fig. \ref{fig:polarization}a presents the slow Rabi oscillation between \ket {52,i} and \ket {52,i'} due to the weak residual $\sigma^-$ component. A fit (solid red line) yields a Rabi oscillation frequency $\Omega_{ii'}^-/2\pi= 107 \pm 25\,\kilo\hertz$. Fig. \ref{fig:polarization}b presents the Autler-Townes doublet induced by the strong $\sigma^+$ component probed on the  \ket {52,i} to \ket{51,n_1=0,m=3} mw transition. We get $\Omega_{ii'}^+/2\pi=\sqrt 2 d_{ii'}E_{\urf}^+/h =  30.0\,\mega\hertz$. The polarization purity is thus remarkably large, with $E_{\urf}^-/((E_{\urf}^+)^2+(E_{\urf}^-)^2)^{1/2} = 0.36\pm 0.08\,\%$. It is consistent with the spatial inhomogeneity of the rf field over the extension of the atomic sample (about one mm) calculated using the SIMION software package.

\begin{figure}[t]
\centering
\includegraphics[width=0.45 \textwidth]{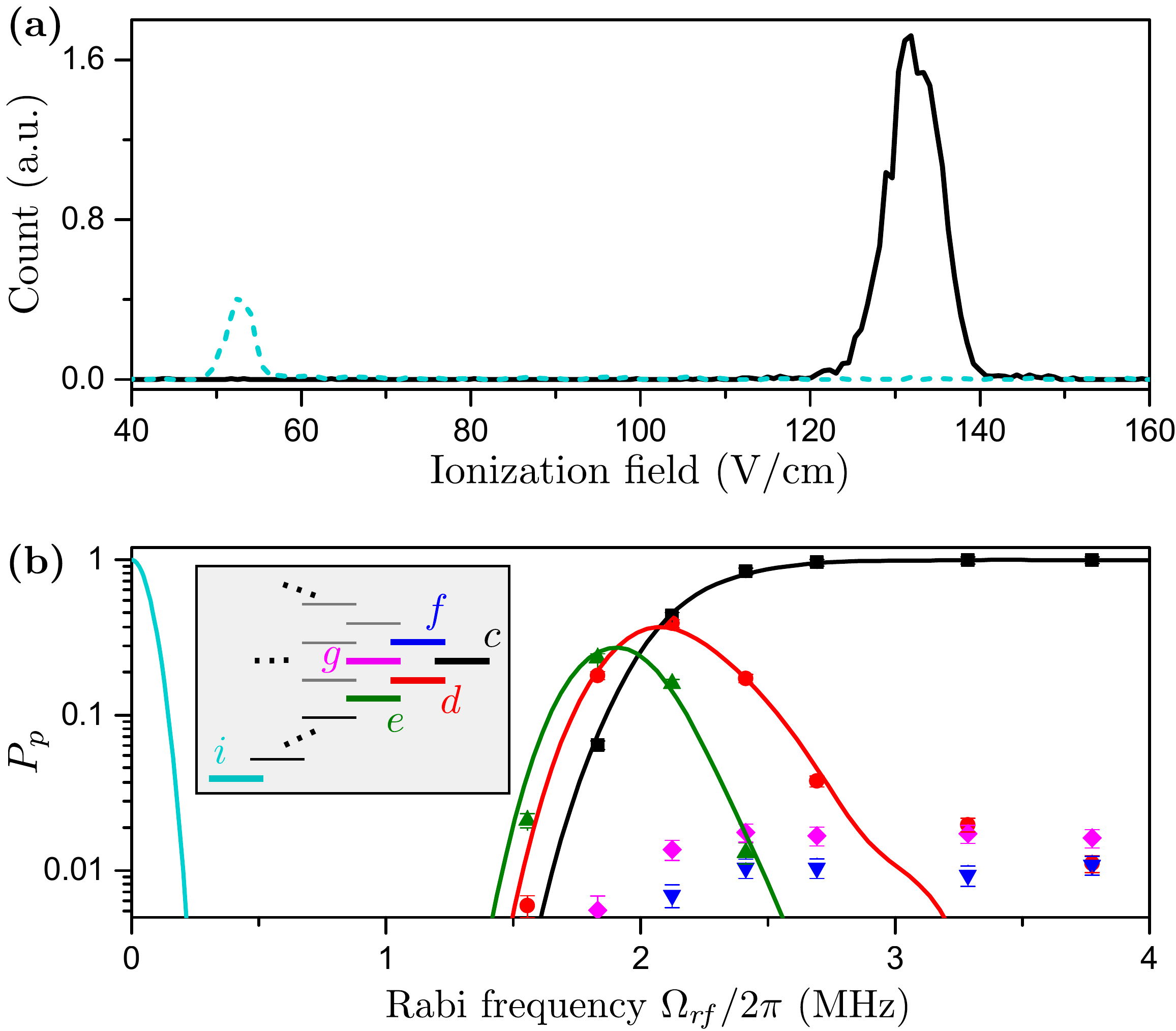}
\caption{(color online) Adiabatic passage in the $n=51$ manifold. (a) Field ionization spectra (signal versus the field applied in the ionization detector) of the initial state $\ket{51,i}$ (dashed cyan line) and at the end of the adiabatic passage sequence (solid black line). (b) Populations $P_p$ for $p=i,c,d,e,f,g$ of the states $\ket{51,p}$ (color code for the levels identification in the inset) as a function of the Rabi frequency $\Omega_{\urf}/2\pi$. Dots are experimental with statistical error bars. The solid lines are the result of a numerical simulation of the full Hamiltonian evolution.}
\label{fig:pa}
\end{figure}

As a first test, we use this optimized 230 MHz rf field to prepare the CRL by the standard adiabatic transfer method~\cite{Nussenzveig1993}. To allow us to use calibrated mw probes \cite{SI}, we perform this experiment in the 51 manifold. We initially prepare the atoms in $\ket{51,i}$ and set the electric field to $F_0=2.45\,\volt\per\centi\meter$ so that $\delta=\omega_{51}-\omega_{\urf}=2\pi\cdot10\,\mega\hertz$. The rf field power is ramped up in $1\,\micro\second$. The electric field $F_0$ is then linearly decreased to $2.24\,\volt\per\centi\meter$ ($\delta=-2\pi\cdot 10\,\mega\hertz$) in $1.5\,\micro\second$, so that $\omega_{51}$ crosses the resonance. The rf is finally switched off in $1\,\micro\second$. We record the number of detected atoms as a function of the ionization field (Fig. \ref{fig:pa}a). The cyan dashed line presents the ionization signal of the initial $\ket{51,i}$ level. The black line shows the signal when the adiabatic passage is performed. It is now centered around the ionization field of the CRL $\ket{51,c}$. From the residual number of atom detected at the ionization field of the $\ket{51,i}$ level, we estimate that the transfer efficiency is larger than 98\%. The difference of height between the cyan and black signals provides the relative detection efficiency, $\eta_0=0.23$, between the $\ket{51,i}$ and $\ket{51,c}$ levels. This low value is mainly due to the difference of lifetimes between the levels. 

We investigate the adiabatic transfer as a function of $\Omega_{\urf}$. The populations of the levels $c, d, e, f, g$ (points in  Fig.~\ref{fig:pa} with statistical error bars) are measured with calibrated mw probes \cite{SI} as a function of the applied rf amplitude, proportional to $E^+_{\urf}$. The conversion factor is determined from the measurement of $\Omega_{ii'}^+$ (Fig.~\ref{fig:polarization}). The data are in excellent agreement with the result of a numerical simulation of the ideal evolution in a pure $\sigma^+$ rf field (solid lines). For  $\Omega_{\urf} > 3$ MHz, the population of the CRL is remarkably close to one, the total population of the levels $d, e, f, g$ being at most 5\% (note that a part of this spurious population is overestimated due to mw probe imperfections). This is another demonstration of the excellent rf polarization control achieved here.

\begin{figure*}[tb]
\centering
\includegraphics[width= 0.8\textwidth]{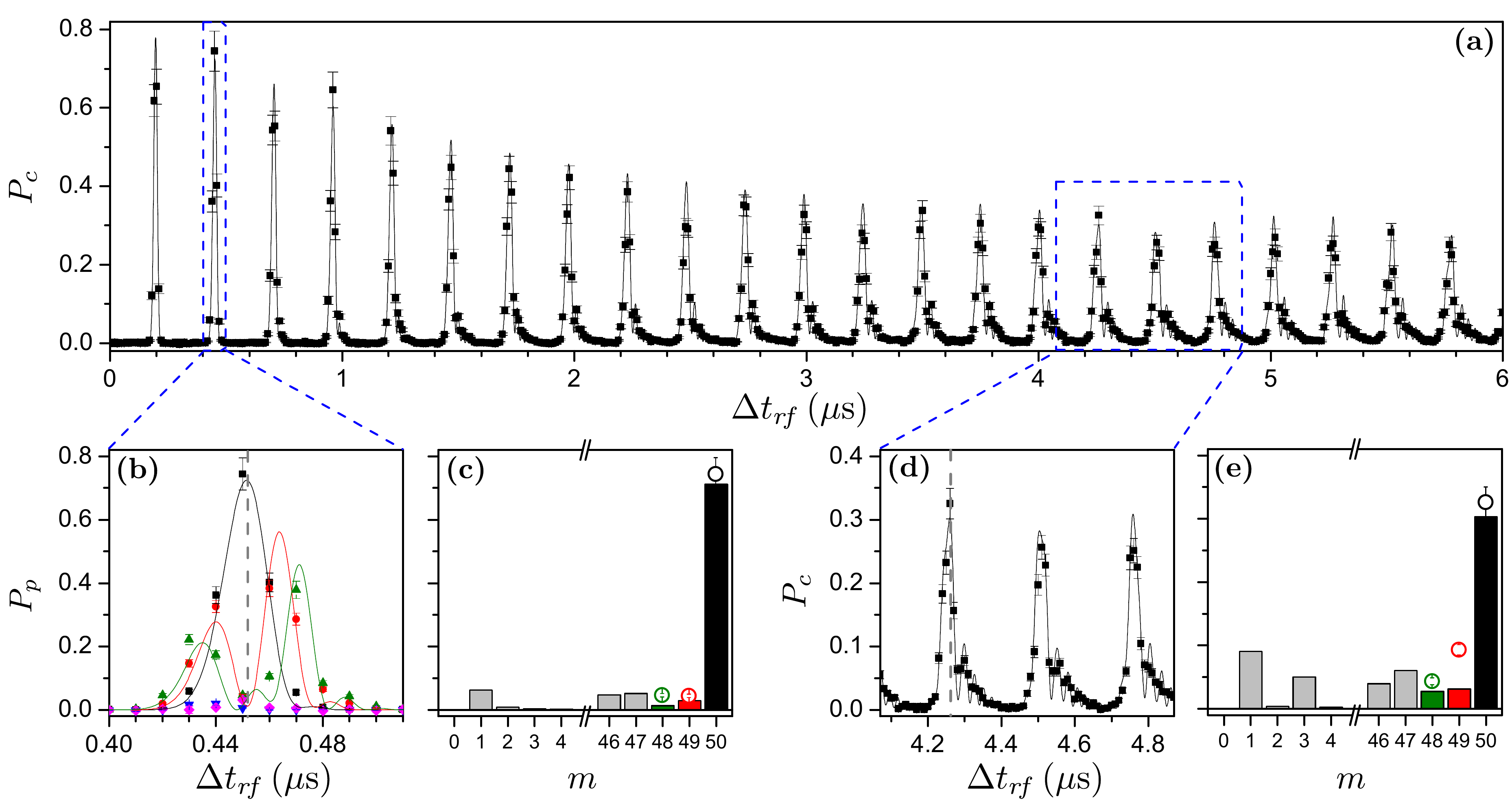}
\caption{(color online) Dynamics of the atomic state for a rf pulse duration $\Delta t_{\urf}$. \textbf{(a)} Long timescale evolution of the CRL population, $P_c$ (dots). The solid line is the result of a numerical simulation, with $\Omega_{\urf}/2\pi=3.52\,\rm{MHz}$ and a rf pulse shortened by $68\,\nano\second$ to model the rise and fall times of the driving electronics. (b) Populations $P_p$ for $p=c,d,e,f,g$ around the second maximum of $P_c$  (same color conventions as in Fig.\ref{fig:polarization}). Dots are experimental and solid lines are the results of the simulation.  (c) Calculated (bars) and measured (open circles) populations $P_p$ at the top of the second CRL population maximum. (d) Enlargement of (a) around three late CRL population maxima. (e) Calculated (bars) and measured (open circles) populations $P_p$ at the top of the first CRL population maximum in (d). For all panels, the experimental error bars correspond to the statistical fluctuations.}
\label{fig:rabi-oscillation}
\end{figure*}

We now investigate the circular state preparation in the resonant rf  regime, which provides the coherent transfer between a low-$\ell$ state and the CRL. The field $F_0$ is adjusted to $2.35\,\volt\per\centi\meter$ corresponding to $\omega_{51}\approx\omega_{\urf}$. We select a fixed rf amplitude. We scan the pulse duration $\Delta t_{\urf}$  and measure the state populations as above. The circular state population $P_c$ (dots in Fig. \ref{fig:rabi-oscillation}a with statistical error bars) exhibits periodic peaks. They reveal more than 20 full Rabi oscillations over 6 $\mu$s, demonstrating the coherence of the atomic evolution. A fit of these data with a numerical simulation of the ideal Hamiltonian evolution in a pure $\sigma^+$ rf field (solid line) provides a calibration of the rf Rabi frequency $\Omega_{\urf} = 3.52$ MHz and of the effective rf pulse duration, which is slightly different ($\approx 70\,\rm{ns}$) from $\Delta t_{\urf}$ due to the finite rise and fall times of the electronic drive.

On Fig.~\ref{fig:rabi-oscillation}b  we plot the populations $P_p$ for $p=i,c,d,e,f,g$ around the second CRL population maximum. We observe for $P_d$ and $P_e$ the theoretically expected double peak structure, typical of a multi-level Rabi oscillation. Fig.~\ref{fig:rabi-oscillation}c presents the calculated (bars) and measured (open circles) population at $\Delta t_{\urf}=0.45\,\rm{\mu s}$. The deviations w.r.t the hydrogen atom model become more important with time, since the quantum defects affect the evolution each time the atom returns to low-$m$ states. Fig~\ref{fig:rabi-oscillation}d shows an enlargement of the evolution at long times (dots) together with the numerical prediction (solid line) and Fig~\ref{fig:rabi-oscillation}e the calculated (bars) and measured (open circles) populations at $\Delta t_{\urf}=4.26\,\rm{\mu s}$.  
The agreement between theory and experiment remains excellent up to large time intervals $\Delta t_{\urf}$, demonstrating the coherence of the oscillation over long time scales.

This experiment shows that it is possible to perform an efficient coherent Rabi oscillation from the low-$m$ state \ket{n,i} to the CRL \ket{n,c} within $200\,\nano\second$.  The limited transfer rate ($\simeq 80$\%) is mainly explained by the the absence of the $m=0,1$ states in the angular momentum ladder due to their large quantum defects and to the residual quantum defect of $m=2$. These residual imperfections could be reduced by tailoring the amplitude and the frequency of the rf field via optimal control \cite{Palao2013} .

We envision a phase gate between an optical and a mw photon, along the lines proposed in~\cite{Pritchard2014} for low-angular momentum Rydberg states. It relies on a dense ensemble of ground state $^{87}$Rb atoms prepared in their $F=1$ hyperfine ground state, and held by a laser dipole trap inside a high-quality open Fabry Perot mw resonator~\cite{Raimond2001a} tuned to the transition between $\ket{50,c}$ and $\ket{51,c}$. An incoming optical photon is transformed into a collective excitation in the $F'=2$ ground state~\cite{Duan2001}. A laser pulse coherently transfers this excitation into the $\ket{50,i}$ Rydberg level, immediately cast onto the $50$ circular state. The atom is then coupled to the cavity for the duration of a resonant $2\pi$ Rabi rotation on the 50 to 51 transition. If the cavity contains one mw photon, this results in a conditional $\pi$ phase shift for the atom \cite{Nogues1999}. This $\pi$ shift is brought back to the $F'$ ground state by the time-reversed excitation process and converted back into an optical photon. All the components of this innovative hybrid photon-photon quantum gate have now been tested separately, including the transfer between the low-$\ell$ and the CRL. Assembling these components would bridge the gap between optical quantum communication and quantum information manipulations based on mw cavity and circuit quantum electrodynamics.

\begin{acknowledgements}
We acknowledge funding by the EU under the ERC projet `DECLIC' (Project ID: 246932) and the RIA project `RYSQ' (Project ID: 640378).
\end{acknowledgements}

\bibliographystyle{h-physrev}

\markboth{Bibliographie}{}
\nocite{}
\bibliography{biblio_circu}
\newpage

\section{Supplemental Material}

We present here the procedure that we followed to optimize the polarization of the rf created by the ring electrodes, and the calibration of the probes that we used to measure the population of the Stark levels $\ket{51,p}$ ($p= c,d,e,f,g$).

\subsection{A. Polarization optimization procedure}

\renewcommand\refname{Supp.Mat.A}
\label{appA}

In order to generate a high-purity $\sigma^+$ polarized rf field we optimize the phases and amplitudes $V_i$ of the rf drives applied on each of the four ring electrodes ($i$ labeling the electrodes from 1 to 4, 1 and 3 on the one hand and 2 and 4 on the other being diametrically opposite). The final adjustment should minimize the $\sigma^-$ component and produce a much larger $\sigma^+$  field with a good spatial homogeneity around the centre of the structure. This adjustment must be impervious to the imperfections due to the complex standing waves in the transmission lines,  to the cross-talks between the capacitively-coupled electrodes and  to the imperfections of the geometry.

The $\sigma^\pm$ components are measured as explained in the main text. The Rabi frequencies are measured using either the spectroscopy of the Autler Townes doublet induced on a probe mw transition or the direct temporal measurement of the induced Rabi oscillation.

We first apply the rf on only one electrode and measure the amplitude of the $\sigma^+$ field it generates when all the others are grounded. We then adjust $V_3$ so that electrode 3 generates a $\sigma^+$ field with the same amplitude as electrode 1, and $V_4$ so that electrode 4 generates a $\sigma^+$ field with the same amplitude as electrode 2. We finally set the relative phase between 1 and 3 (respectively 2 and 4), in order to maximize the amplitude of the $\sigma^+$ field generated when electrodes 1 and 3 (2 and 4) are driven simultaneously.

We now cancel the total $\sigma^-$ component. We first measure the amplitude of the $\sigma^-$ component created by each pair of electrodes (pair 1-3 or pair 2-4) separately. We scale $V_2$ and $V_4$ together so that pair 2-4 generates the same $\sigma^-$ amplitude as pair 1-3. Finally, driving the four electrodes, we set the relative phases between the two pairs to minimize the global $\sigma^-$ amplitude. 

From this initial setting, the $\sigma^+$ field amplitude is varied by scaling with the same factor all the amplitudes applied on the four electrodes by means of identical variables attenuators driven by a common signal.

\bigskip
\subsection{B. Calibration of the microwave probes efficiency}
\label{appB}

In order to measure the populations of the Stark levels $\ket{51,p}$ ($p= c,d,e,f,g$) we use mw probes tuned to the two-photon transitions from $\ket{51,p}$ to $\ket{49,p}$, which are resolved in the applied electric field of 3,8 V/cm. For each transition, we perform a $\pi$ pulse ideally transferring all the population of \ket{51,p} into \ket{49,p}. In practice each $\pi$ pulse has a limited efficiency. Moreover, the detection efficiency of levels \ket{51,p} and \ket{49,p} is different. We denote by $\eta_p$ the overall transfer efficiency between \ket{49,p} and \ket{51,p}. We estimate it by preparing selectively $\ket{51,p}$ and measuring the fraction of atoms detected at the ionization threshold of \ket{49,p} after a probe pulse.

We prepare $\ket{51,p}$ by first exciting \ket{52,i} or \ket{52,j} with a laser pulse. We then transfer the atoms into the level $\ket{51,q}$  ($q= i, j, k, l$) using a mw pulse. We finally apply a rf adiabatic passage, which ideally maps $\ket{51,q}$ with $q= i, j, k, l$ into $\ket{51,p}$ with $p=c, d, e, f$ respectively. 

Due to imperfections in the preparation process, the states neighboring \ket{51,p} are also slightly populated, which leads to underestimate $\eta_p$. For levels $c,d$ and $e$, theses spurious population are small and the \ket{51,d} and \ket{51,e} populations in figures~3 and~4 are slightly overestimated. The preparation of $f$ leads to comparable populations in $f$ and $g$. We use them both to get an underestimated value of $\eta_f$ and $\eta_g$. 

The probability $P_p$ to end-up in the state \ket{51,p} is then
$$ P_p = \frac {\eta_0}{\eta_p}  \frac {N(49,p)}{N(51,i)} \,,$$
where $N(49,p)$ is the number of atom detected in \ket{49,p} after  the mw probe pulse, and $N(51,i)$ is the initial population in the laser-excited \ket{51,i}.

\end{document}